\documentclass[twocolumn,showpacs,showkeys,preprintnumbers,amssymb,aps,superscriptaddress,prb]{revtex4}
\usepackage{graphicx}
\usepackage{dcolumn}
\usepackage{bm}

\usepackage{epsfig}

\begin{document}

\preprint{APS/123-QED}

\title{Unconventional quantum ordered and disordered states in the highly frustrated \\ spin-1/2 Ising-Heisenberg model on triangles-in-triangles lattices}

\author{Jana \v{C}is\'{a}rov\'{a}}
\email{jana.kissova@student.upjs.sk}
\affiliation{Department of Theoretical Physics and Astrophysics, Faculty of Science, 
P. J. \v{S}af\'{a}rik University, Park Angelinum 9, 040 01, Ko\v{s}ice, Slovak republic}
\affiliation{Institute of Theoretical Physics, Ecole Polytechnique F\'{e}d\'{e}rale de Lausanne, CH-1015 Lausanne, Switzerland}
\author{Jozef Stre\v{c}ka}
\email{jozef.strecka@upjs.sk}
\affiliation{Department of Theoretical Physics and Astrophysics, Faculty of Science, 
P. J. \v{S}af\'{a}rik University, Park Angelinum 9, 040 01, Ko\v{s}ice, Slovak republic}

\date{\today}

\begin{abstract}
The spin-1/2 Ising-Heisenberg model on two geometrically related triangles-in-triangles lattices is exactly solved through the generalized star-triangle transformation, which establishes a rigorous mapping correspondence with the effective spin-1/2 Ising model on a triangular lattice. Basic thermodynamic quantities were exactly calculated within  this rigorous mapping method along with the ground-state and finite-temperature phase diagrams. Apart from the classical ferromagnetic phase, both investigated models exhibit several unconventional quantum ordered and disordered ground states. A mutual competition between two ferromagnetic interactions of basically different character generically leads to the emergence of a quantum ferromagnetic phase, in which a symmetric quantum superposition of three up-up-down states of the Heisenberg trimers accompanies a perfect alignment of all Ising spins. In the highly frustrated regime, we have either found the disordered quantum paramagnetic phase with a substantial residual entropy or a similar but spontaneously ordered phase with a nontrivial criticality at finite temperatures. The latter quantum ground state exhibits a striking coexistence of imperfect spontaneous order with partial disorder, which is evidenced by a quantum reduction of the spontaneous magnetization of Heisenberg spins that indirectly causes a small reduction of the spontaneous magnetization of otherwise classical Ising spins. 
\end{abstract}

\pacs{05.50.+q, 64.60.F-, 75.10.Jm, 75.30.Kz, 75.40.Cx}

\keywords{Ising-Heisenberg model, geometric spin frustration, triangles-in-triangles lattices}

\maketitle

\section{Introduction}

Frustrated two-dimensional (2D) quantum spin models belong to the most challenging issues to deal with in the modern statistical and condensed matter physics.\cite{lhui02,misg04,rich04,lacr11} A considerable interest in investigating 2D geometrically frustrated spin models is closely connected with an existence of numerous layered insulating magnetic materials,\cite{lacr11,busc01,gree01,harr04} which can be characterized by incapability of spins to simultaneously satisfy all pair spin-spin interactions. In general, the magnetism of the vast majority of insulating materials is well captured by the quantum Heisenberg model and its various extensions.\cite{jong74} It should be mentioned, however, that a mutual interplay between the geometric frustration and quantum fluctuations is not only responsible for an extraordinary diverse magnetic behavior, but is also the main cause of a mathematical intractability of geometrically frustrated Heisenberg models with exception of a few prominent exactly solved cases.\cite{miya11} On the other hand, the classical Ising model is still exactly solvable on several frustrated 2D lattices and it may thus provide a valuable insight into cooperative phenomena originating from the spin-frustration effect.\cite{lieb86,diep04} The most limiting drawback of the Ising model unfortunately lies in the lack of appropriate experimental realizations which are rather scarce.\cite{jong74,wolf00} Among the most striking phenomena emerging in geometrically frustrated classical and quantum spin systems one could mention the existence of reentrant phase transitions,\cite{lieb86,diep04} the quantum reduction of the magnetization,\cite{barn91,mano91} the presence of magnetization jumps and intermediate plateaus in the magnetization process,\cite{rich04,schu04,hone04} "order-from-disorder" effect,\cite{vill77,vill80} the appearance of several unusual spin-liquid ground states,\cite{lhui02,misg04,lacr11} the enhanced magnetocaloric effect,\cite{zhit03,hone06,derz06}, the localized magnons in a close vicinity of saturation field,\cite{schu02,zhit04,derz04,zhit05} etc.

Despite a great effort in this research area, the classical and quantum frustrated spin models defined on 2D "triangles-in-triangles" (TIT) lattices have received much less attention so far. TIT lattices generally consists of smaller triangular entities embedded in either some or all triangular cells of 2D triangle-based lattices such as triangular or kagom\'e lattice. The sites of original lattice might be thus called as nodal lattice sites, while the sites of smaller triangular entities may be viewed as decorating lattice sites. In this way, one may consider various geometrically frustrated spin models on different but geometrically related TIT lattices. If all lattice sites are occupied by the Ising spins, one considers the classical Ising model whose magnetic behavior is not affected by quantum fluctuations at all. If all lattice sites are occupied by quantum Heisenberg spins, one turns to the quantum Heisenberg model that is basically affected through quantum effects. If the nodal lattice sites are occupied by the Ising spins and the decorating sites by the quantum Heisenberg ones, one considers the hybrid Ising-Heisenberg model affected through local quantum fluctuations only.

Up to now, there exist only a few theoretical studies of classical and quantum spin models on one very special example of TIT lattice called as the triangulated kagom\'e lattice\cite{zhen05,loh08,stre08,yao08,stre09,nato97,stre07,isod11,isod12,rich12} (see for instance Fig.~1 of Ref. \onlinecite{stre08} for the pictorial representation of this lattice). A particular interest in studying the role of geometric frustration in the triangulated kagom\'e lattice has been stimulated by the family of polymeric coordination compounds Cu$_9$X$_2$(cpa)$_6$$\cdot$nH$_2$O (X=F,Cl,Br and cpa=carboxypentonic acid), which provides
an experimental realization of highly frustrated magnetic materials with a remarkably high frustration ratio ($>130$) and possibly spin-liquid ground state.\cite{gonz93,maru94,atec95,okub98,meka98,meka01} In addition, there are also experimental indications of one-third plateau in the low-temperature magnetization process,\cite{meka98,meka01} while the temperature dependence of the susceptibility implies two basically different exchange interactions in this intriguing series of magnetic compounds.\cite{maru94,atec95,okub98,meka98,meka01} A stronger antiferromagnetic coupling has been assigned to the exchange interaction inside the smaller triangles and a possibly much weaker ferromagnetic coupling to the exchange interaction within the larger triangles of the triangulated kagom\'e lattice.\cite{meka98,meka01} Electron spin resonance (ESR) measurements gave the almost isotropic g-value and this result is taken to mean that Cu$_9$X$_2$(cpa)$_6$$\cdot$nH$_2$O compounds afford a suitable experimental realization of the isotropic Heisenberg model.\cite{okub98,meka01} Of course, the quantum Heisenberg model on the triangulated kagom\'e lattice cannot be rigorously solved and thus, this model has been dealt with the help of different approximate perturbation approaches\cite{nato97,stre07} and numerical exact diagonalization.\cite{isod11,isod12,rich12} However, the substantial difference between the exchange interactions within the smaller and larger triangles of the triangulated kagom\'e lattice provides a plausible ground for introducing the simplified Ising-Heisenberg model, which correctly reproduces the stronger Heisenberg coupling inside the smaller triangles and merely approximates the much weaker exchange interaction inside the larger triangles by the Ising coupling. A comparison between the exact solutions of the Ising \cite{zhen05,loh08} and Ising-Heisenberg \cite{stre08,yao08,stre09} models on triangulated kagom\'e lattice reveals that a presence of local quantum fluctuations leads to a marked decrease in the residual entropy of the quantum paramagnetic state of the latter classical--quantum model. 

On the other hand, there is a lack of theoretical studies concerning with the classical and quantum spin models defined on other geometrically frustrated TIT lattices, which could be derived for instance from a simple triangular lattice. The main goal of the present work is to fill in this gap, whereas the main emphasis will be laid on a comparative study of the Ising-Heisenberg models defined on two different but geometrically related TIT lattices each of them descended from a simple triangular lattice. It will be demonstrated that two different magnetic architectures, even though they are geometrically closely related, will cause a different level of local quantum fluctuations and thus fundamentally change the magnetic behavior of the considered spin models especially in the highly frustrated regime. It actually turns out that the ground state of the first considered model in the highly frustrated regime is disordered in contrast to the partially ordered and partially disordered ground state of the second model, which is characterized by the obvious quantum reduction of the spontaneous magnetization.      

The organization of this paper is as follows. Both investigated models are introduced in Section \ref{sec:model} along with basic steps of their exact analytical treatment. 
The most interesting results for the ground state, order parameters and overall thermodynamics are then discussed in Section \ref{sec:result}. Finally, several concluding remarks are drawn in Section \ref{sec:conclusion}. 

\section{Model and solution}
\label{sec:model}

\begin{figure*}[t]
\vspace{-0.2cm}
\includegraphics[width=14.0cm]{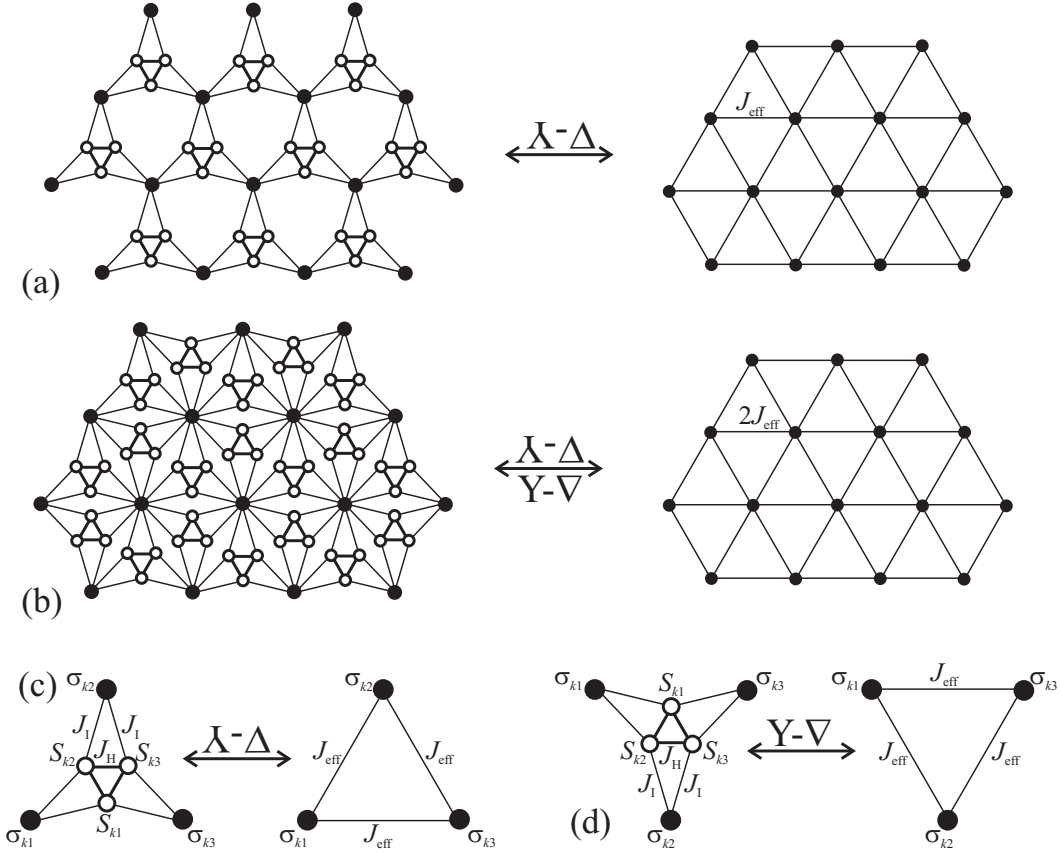}
\vspace{-0.2cm}
\caption{(a)-(b) The spin-1/2 Ising-Heisenberg model on two different but geometrically related TIT lattices and its rigorous mapping to the spin-1/2 Ising model on a triangular lattice. The full circles denote lattice positions of the Ising spins $\sigma=1/2$ and the empty ones lattice positions of the Heisenberg spins $S=1/2$; (c)-(d) Schematic representation of two equivalent star-triangle transformations used for establishing a precise mapping correspondence between both the models.}
\label{fig1}
\end{figure*}

Let us introduce the spin-1/2 Ising-Heisenberg model on two different TIT lattices, which are schematically depicted on the left-hand-side of Fig.~\ref{fig1}(a)-(b). As illustrated, the Ising spins $\sigma=1/2$ placed at nodal lattice sites of a simple triangular lattice (full circles) are mutually inter-connected through smaller triangles of the quantum Heisenberg spins $S=1/2$ (empty circles), which are either placed into each up-pointing triangle (Fig.~\ref{fig1}(a)) or into each triangle (Fig.~\ref{fig1}(b)) of the underlying triangular lattice.  From this perspective, the smaller triangles of three quantum Heisenberg spins (Heisenberg trimers) are embedded in the larger triangles of the classical Ising spins and hence, both investigated magnetic structures fall into the class of TIT lattices. The total Hamiltonian of the spin-1/2 Ising-Heisenberg model on both aforedescribed TIT lattices can be defined as
\begin{eqnarray}
\hat{\cal H} \!=\! -J_{\rm H} \!\!\! \sum_{<i,j>}^{\circ\!-\!\circ} \!\!\! \left[ \Delta \! \left( \hat{S}_i^x \hat{S}_{j}^x \!+\! \hat{S}_{i}^y \hat{S}_{j}^y \right)
 \!+\! \hat{S}_i^z \hat{S}_{j}^z \right] \!-\! J_{\rm I} \!\!\! \sum_{<k,l>}^{\circ\!-\!\bullet} \!\!\! \hat{\sigma}_k^z \hat{S}_l^z\!,  
\label{Htot} 
\end{eqnarray}
where $\hat{\sigma}_{k}^{z}$ and $\hat{S}_{i}^{\alpha}$ ($\alpha=x,y,z$) denote spatial components of the spin-1/2 operator of the Ising and Heisenberg spins, respectively.  The parameter $J_{\rm H}$ stands for the XXZ Heisenberg interaction between the nearest-neighbor Heisenberg spins from the small inner triangles, $\Delta$ is a spatial anisotropy in this interaction and the parameter $J_{\rm I}$ represents the Ising-type interaction between the nearest-neighbor Ising and Heisenberg spins. 

For further convenience, let us rewrite the total Hamiltonian as a sum over the six-spin cluster Hamiltonians schematically illustrated in Fig.~\ref{fig1}(c)-(d)
\begin{eqnarray}
\hat{\cal H} = \sum_{k=1}^{\gamma N} {\hat{\cal H}_k}. 
\label{0}
\end{eqnarray}
Here, $N$ stands for the total number of the Ising spins and $\gamma N$ denotes the total number of six-spin clusters, i.e. $\gamma = 1$ for the TIT lattice drawn in Fig.~\ref{fig1}(a) and $\gamma = 2$ for the TIT lattice shown in Fig.~\ref{fig1}(b). It is of principal importance that all the interaction terms including the 
three Heisenberg spins from the $k$th inner triangle belong to the same cluster Hamiltonian $\hat{\cal H}_k$ 
\begin{eqnarray}
\hat{\cal H}_{k} \!\!\!&=&\!\!\! -J_{\rm H} \sum_{i=1}^{3} [\Delta (\hat{S}_{k,i}^x \hat{S}_{k,i+1}^x + \hat{S}_{k,i}^y\hat{S}_{k,i+1}^y)
+ \hat{S}_{k,i}^z \hat{S}_{k,i+1}^z] \nonumber \\
 \!\!\!&-&\!\!\!  J_{\rm I} \sum_{i=1}^{3} \hat{\sigma}_{k,i}^z (\hat{S}_{k,i}^z  + \hat{S}_{k,i+1}^z), \,\,\, \qquad \qquad (\hat{S}_{k,4}^{\alpha} \equiv \hat{S}_{k,1}^{\alpha}) \nonumber \\
\label{1}
\end{eqnarray}
which in turn ensures a validity of the standard commutation relation $[\hat{\cal H}_i, \hat{\cal H}_j] = 0$ between different cluster Hamiltonians $i \neq j$. Owing to this fact, the total partition function can be partially factorized into a product of cluster partition functions and one may trace out spin degrees of freedom of different Heisenberg trimers independently of each other according to the relation
\begin{eqnarray}
{\cal Z} = \sum_{\{ \sigma_i \}} \prod_{k=1}^{\gamma N} {\rm Tr}_k \exp(-\beta \hat{\cal H}_k) = \sum_{\{ \sigma_i \}} \prod_{k=1}^{\gamma N} {\cal Z}_k,
\label{2}
\end{eqnarray}
where $\beta = 1/(k_{\rm B} T)$, $k_{\rm B}$ is Boltzmann's constant, $T$ is the absolute temperature, the symbol $\sum_{\{ \sigma_i \}}$ denotes a summation over spin states of all the Ising spins and the symbol ${\rm Tr}_k$ stands for a trace over spin degrees of freedom of the $k$th Heisenberg trimer. After tracing out the spin degrees of freedom of the $k$th Heisenberg trimer, the cluster partition function ${\cal Z}_k$ will solely depend on the three nodal Ising spins $\sigma_{k1}$, $\sigma_{k2}$, and $\sigma_{k3}$ attached to the $k$th Heisenberg trimer. Moreover, one may take advantage of the generalized star-triangle transformation \cite{fish59,syoz72,roja09,stre10} 
\begin{eqnarray}
{\cal Z}_k \!\!\!\!\!&&\!\!\!\!\! (\sigma_{k1}^z, \sigma_{k2}^z, \sigma_{k3}^z) = {\rm Tr}_k \exp(-\beta \hat{\cal H}_k) \nonumber  \\
 \!\!\!&=&\!\!\! A \exp \left[\beta J_{\rm eff} \left(\sigma_{k1}^z \sigma_{k2}^z + \sigma_{k2}^z \sigma_{k3}^z + \sigma_{k3}^z \sigma_{k1}^z \right) \right], 
 \label{stt} 
\end{eqnarray}
which can be used in order to replace this effective Boltzmann's factor by the equivalent expression depending on the three nodal Ising spins only. 
The above mapping transformation generally represents a set of eight algebraic equations, which can be obtained from Eq.~(\ref{stt}) by substituting all available spin configurations of three Ising spins involved therein. In an absence of the external magnetic field, one actually gets just two independent equations; the first equation 
for two uniform configurations with the three equally aligned Ising spins 
\begin{eqnarray}
V_1 \!\!\!&\equiv&\!\!\! {\cal Z}_k (\pm 1/2, \pm 1/2, \pm1/2) \nonumber \\
 \!\!\!&=&\!\!\! 2 \exp \left( \frac{3}{4} \beta J_{\rm H} \right) \cosh \left( \frac{3}{2} \beta J_{\rm I} \right) \nonumber \\
 \!\!\!&+&\!\!\! 2 \exp \left[ -\frac{1}{4} \beta J_{\rm H}(1 - 4\Delta) \right] \cosh \left( \frac{1}{2} \beta J_{\rm I}\right) \nonumber \\
 \!\!\!&+&\!\!\!  4 \exp \left[ -\frac{1}{4} \beta J_{\rm H}(1 + 2 \Delta) \right] \cosh \left( \frac{1}{2}\beta J_{\rm I} \right) \nonumber \\
 \!\!\!&=&\!\!\! A \exp \left(\frac{3}{4} \beta J_{\rm eff}\right),
\label{3}
\end{eqnarray}
and the second equation for six non-uniform configurations with one Ising spin pointing in opposite with respect to the other two 
\begin{eqnarray}
V_2 \!\!\!&\equiv&\!\!\! {\cal Z}_{k} (\pm 1/2, \pm 1/2, \mp 1/2) = {\cal Z}_{k} (\pm 1/2, \mp 1/2, \pm 1/2) \nonumber \\
\!\!\!&=&\!\!\! {\cal Z}_{k} (\mp 1/2, \pm 1/2, \pm 1/2) \nonumber \\
\!\!\!&=&\!\!\! 2 \exp \left( \frac{3}{4}\beta J_{\rm H} \right) \cosh \left( \frac{1}{2} \beta J_{\rm I} \right) \nonumber \\
\!\!\!&+&\!\!\! 2 \exp \left[-\frac{1}{4} \beta J_{\rm H}(1 + 2\Delta)\right]  \cosh \left( \frac{1}{2}\beta J_{\rm I} \right) \nonumber \\
\!\!\!&+&\!\!\!  2 \exp \left[ -\frac{1}{4} \beta J_{\rm H}(1 - \Delta) \right] \cosh \left( \frac{1}{2}\beta Q^{+} \right) \nonumber \\
\!\!\!&+&\!\!\! 2 \exp \left[- \frac{1}{4}\beta J_{\rm H}(1 - \Delta)\right] \cosh \left( \frac{1}{2}\beta Q^{-} \right) \nonumber \\
\!\!\!&=&\!\!\! A \exp \left(-\frac{1}{4} \beta J_{\rm eff}\right).
\label{4}
\end{eqnarray}
For abbreviation, we have introduced here two auxiliary functions $Q^{\pm}$ defined as follows
\begin{eqnarray}
Q^{\pm} = \sqrt{\left(\frac{J_{\rm H} \Delta}{2} \pm J_{\rm I} \right)^{2} + 2 (J_{\rm H} \Delta)^{2}}.
\label{5}
\end{eqnarray}
It is noteworthy that two algebraic equations (\ref{3})--(\ref{4}) unambiguously determine 
yet unspecified mapping parameters $A$ and $J_{\rm eff}$
\begin{eqnarray}
A = {\left( V_1 V_2^3 \right)}^{\frac{1}{4}}, \qquad \beta J_{\rm eff} = \ln \left( \frac{V_1}{V_2}\right),
\label{par}
\end{eqnarray}
which ensure a general validity of the star-triangle transformation (\ref{stt}) on assumption that 
the mapping parameters $A$ and $J_{\rm eff}$ are given by Eqs.~(\ref{3})--(\ref{par}).

Substituting the generalized star-triangle transformation (\ref{stt}) into the relation (\ref{2}) one easily gets a rigorous mapping correspondence 
between the partition function ${\cal Z}$ of the spin-1/2 Ising-Heisenberg model on the TIT lattice and respectively, the partition function ${\cal Z}_{\rm IM}$ 
of the spin-1/2 Ising model on a simple triangular lattice   
\begin{eqnarray}
{\cal Z} (\beta, J_{\rm H}, J_{\rm I}, \Delta) = A^{\gamma N} {\cal Z}_{\rm IM} (\beta, \gamma J_{\rm eff}), 
\label{Z}
\end{eqnarray}
which is unambiguously given by the Hamiltonian with the effective nearest-neighbor interaction $\gamma J_{\rm eff}$
\begin{eqnarray}
{\cal H}_{\rm IM} = - \gamma J_{\rm eff} \sum_{<i,j>} \sigma_i^z \sigma_j^z. 
\label{him}
\end{eqnarray}
Despite the different spatial symmetries, the Ising-Heisenberg model on the two distinct TIT lattices can be rigorously mapped to the same effective Ising model 
on a simple triangular lattice, since the generalized star-triangle transformation (\ref{stt}) is performed locally for the same smallest commuting spin cluster 
in both the models (see Fig. \ref{fig1}(c)-(d)). As a matter of fact, the most crucial difference between both the mappings 
lies just in a strength of the effective nearest-neighbor coupling $\gamma J_{\rm eff}$, which is twice as large for the TIT lattice shown in Fig. \ref{fig1}(b) 
compared to that displayed in Fig. \ref{fig1}(a). The other mapping parameter $A$ represents just less important multiplicative factor in the mapping relation 
(\ref{Z}) between both partition functions. It is worthwhile to remark, moreover, that the exact result for the partition function of 
the spin-1/2 Ising model on the triangular lattice is well known \cite{hout50,temp50,wann50,domb60} 
\begin{eqnarray}
\ln \left( \frac{{\cal Z}_{\rm IM}}{2} \right) \!\!&=&\!\!  \frac{1}{8 \pi^2} \! \int_{0}^{2 \pi} \!\!\! \int_{0}^{2 \pi} \ln \biggl[C_1 - D_1 \cos \theta 
\nonumber \\  &-& D_1 \cos \phi \!-\! D_1 \cos(\theta + \phi) \biggr] {\rm d} \theta {\rm d} \phi, \nonumber \\
C_1 &=& \cosh^3 \left( \frac{\beta \gamma J_{\rm eff}}{2} \right) + \sinh^3 \left( \frac{\beta \gamma J_{\rm eff}}{2} \right) \nonumber \\ 
D_1 &=& \sinh \left( \frac{\beta \gamma J_{\rm eff}}{2} \right),
\label{PFTr}
\end{eqnarray}
and hence, the partition function of the spin-1/2 Ising-Heisenberg model on the TIT lattices can also be exactly calculated by inserting the exact result (\ref{PFTr}) 
for the partition function ${\cal Z}_{\rm IM}$ of the spin-1/2 Ising model on the triangular lattice with the effective nearest-neighbor interaction 
$\gamma J_{\rm eff}$ into the rigorous mapping relation (\ref{Z}).

At this stage, other basic thermodynamic quantities can also be calculated in a rather straightforward way within the framework of this rigorous mapping method. 
Using the standard relations of thermodynamics and statistical physics, the free energy $F$ of the spin-1/2 Ising-Heisenberg model on the TIT lattices can be expressed 
through the free energy $F_{\rm IM}$ of the corresponding spin-1/2 Ising model on the triangular lattice 
\begin{eqnarray}
F = F_{\rm IM} (\beta, \gamma J_{\rm eff}) - \gamma N k_{\rm B} T \ln A.
\label{F}
\end{eqnarray}
Similarly, the internal energy $U$ of the spin-1/2 Ising-Heisenberg model on the TIT lattices can be connected to the internal energy $U_{\rm IM}$ of the spin-1/2 Ising model on the triangular lattice with the temperature-dependent nearest-neighbor interaction $\gamma J_{\rm eff}$ 
\begin{eqnarray}
U \!\!=\!\! - \frac{\gamma N}{4} \! \left(\frac{W_1}{V_1} \!+\! \frac{3 W_2}{V_2} \right) 
 \!+\! \frac{U_{\rm IM} (\beta, \gamma J_{\rm eff})}{J_{\rm eff}} \! \left(\frac{W_1}{V_1} \!-\! \frac{W_2}{V_2} \right)\!\!.
\label{6}
\end{eqnarray}
For completeness, let us merely quote that the explicit form of the internal energy $U_{\rm IM}$ of the spin-1/2 Ising model on the triangular lattice can be for instance found in Ref.~[\onlinecite{domb60}], while two new functions $W_1$ and $W_2$ entering the established mapping relationship (\ref{6}) for the internal energy are explicitly 
given by the following formulas 
\begin{widetext}
\begin{eqnarray}
W_1 = \frac{\partial V_1}{\partial \beta} &=& \frac{3}{2}\exp \left(\frac{3}{4}\beta J_{\rm H}\right) \left[ J_{\rm H}
 \cosh \left( \frac{3}{2} \beta J_{\rm I} \right) + 2 J_{\rm I} \sinh \left( \frac{3}{2} \beta J_{\rm I} \right) \right] \nonumber \\
&-&  \frac{1}{2} \exp \left[ -\frac{1}{4} \beta J_{\rm H}(1 - 4\Delta) \right]
 \left[ J_{\rm H} (1 - 4 \Delta) \cosh \left( \frac{1}{2}\beta J_{\rm I} \right) - 2 J_{\rm I} \sinh \left( \frac{1}{2}\beta J_{\rm I} \right) \right]
 \nonumber \\
 &-& \exp \left[ -\frac{1}{4} \beta J_{\rm H}(1 + 2 \Delta) \right] \left[  J_{\rm H}
 (1 + 2 \Delta) \cosh \left( \frac{1}{2}\beta J_{\rm I} \right) - 2 J_{\rm I} \sinh \left( \frac{1}{2} \beta J_{\rm I} \right)\right], \nonumber \\
W_2  = \frac{\partial V_2}{\partial \beta} &=& \frac{3}{2} \exp \left( \frac{3}{4}\beta J_{\rm H} \right)
\left[ J_{\rm H} \cosh \left( \frac{1}{2} \beta J_{\rm I} \right) + \frac{2}{3} J_{\rm I} \sinh \left( \frac{1}{2} \beta J_{\rm I} \right) \right] \nonumber \\
 &-& \frac{1}{2} \exp \left[-\frac{1}{4} \beta J_{\rm H}(1 + 2 \Delta) \right]
\left[ J_{\rm H} (1 + 2 \Delta) \cosh \left( \frac{1}{2}\beta J_{\rm I} \right) - 2 J_{\rm I} \sinh \left( \frac{1}{2} \beta J_{\rm I} \right) \right] \nonumber \\
&-&  \frac{1}{2} J_{\rm H} (1 - \Delta) \exp \left[ -\frac{1}{4} \beta J_{\rm H}(1 - \Delta) \right]
 \left[ \cosh \left( \frac{1}{2}\beta Q^{+} \right) + \cosh \left( \frac{1}{2}\beta Q^{-} \right) \right] \nonumber \\ 
&+&  \exp \left[- \frac{1}{4}\beta J_{\rm H}(1 - \Delta) \right]
 \left[ Q^{+} \sinh \left( \frac{1}{2}\beta Q^{+}\right) + Q^{-}\sinh \left( \frac{1}{2}\beta Q^{-} \right) \right]. 
\label{6b}
\end{eqnarray} 
\end{widetext}
Furthermore, one may also readily obtain exact results for the entropy $S$ and specific heat $C$ by making use of the standard thermodynamic relations  
\begin{eqnarray}
S = k_{\rm B} \ln {\cal Z} + \frac{U}{T}, \qquad C = \frac{\partial U}{\partial T},
\label{SC} 
\end{eqnarray}
after substituting the previously derived exact results for the partition function (\ref{Z}) and the internal energy (\ref{6}) into the relations (\ref{SC}). 

Next, let us calculate the spontaneous magnetization of the Ising and Heisenberg spins, respectively. The spontaneous magnetization of the Ising spins can be
simply calculated by exploiting the exact mapping theorems developed by Barry and co-workers \cite{barr88,khat90,barr91,barr95}
\begin{eqnarray}
\langle f_1 (\hat{\sigma}_i^z, \hat{\sigma}_j^z, \ldots, \hat{\sigma}_k^z) \rangle = 
\langle f_1 (\hat{\sigma}_i^z, \hat{\sigma}_j^z, \ldots, \hat{\sigma}_k^z) \rangle_{\rm IM}, \label{si1} 
\end{eqnarray}
according to which the canonical ensemble average $\langle \cdots \rangle$ of any function $f_1$ of the Ising spins in the spin-1/2 Ising-Heisenberg model on the TIT lattice  directly equals the canonical ensemble average $\langle \cdots \rangle_{\rm IM}$ of the same function $f_1$ of the Ising spins in the equivalent spin-1/2 Ising model 
on the triangular lattice. This result is taken to mean that the spontaneous magnetization of the Ising spins $m_{\rm I}$ in the spin-1/2 Ising-Heisenberg model on the TIT lattice can in turn be calculated from the single-site magnetization $m_{\rm IM}$ of the spin-1/2 Ising model on the triangular lattice \cite{pott52} 
\begin{eqnarray}
m_{\rm I} \!\!\!&\equiv&\!\!\! \langle \hat{\sigma}_i^z \rangle = \langle \hat{\sigma}_i^z \rangle_{\rm IM} \equiv m_{\rm IM} (\beta, \gamma J_{\rm eff}), \nonumber \\
m_{\rm IM} \!\!\!&=&\!\!\! \frac{1}{2} \left[1 - \frac{16 z^{6}}{(1 + 3z^2)(1 - z^2)^3}\right]^{\frac{1}{8}}, 
\label{m0}
\end{eqnarray}
with the parameter $z = \exp(- \gamma \beta J_{\rm eff}/2)$. Substituting the explicit form (\ref{par}) of the effective nearest-neighbor interaction $\gamma J_{\rm eff}$ 
into the formula (\ref{m0}) one finally gets the resultant exact expression for the single-site magnetization of the Ising spins in the spin-1/2 Ising-Heisenberg model 
on the TIT lattice
\begin{eqnarray}
m_{\rm I} = \frac{1}{2} \left[1 - \frac{16 V_1^{\gamma} V_2^{3 \gamma}}{\left(V_1^{\gamma} + 3 V_2^{\gamma}\right)\left(V_1^{\gamma} - V_2^{\gamma}\right)^3}\right]^{\frac{1}{8}}, 
\label{mi}
\end{eqnarray}
whereas $\gamma=1$ for the TIT lattice shown in Fig. \ref{fig1}(a) and $\gamma=2$ for the TIT lattice drawn in Fig. \ref{fig1}(b). 

Owing to the commuting character of different cluster Hamiltonians, the ensemble average of any function $f_2$ of the Ising and Heisenberg spins belonging to the $k$th cluster Hamiltonian (\ref{1}) can be calculated from the generalized form of Callen-Suzuki spin identity \cite{call63,suzu65,balc02}
\begin{eqnarray}
\!\!\!\!\!\!&&\!\!\!\!\!\! \langle f_2 
(\hat{S}_{k1}^{\alpha}, \hat{S}_{k2}^{\alpha}, \hat{S}_{k3}^{\alpha}, \hat{\sigma}_{k1}^{z}, \hat{\sigma}_{k2}^{z}, \hat{\sigma}_{k3}^{z}) \rangle =  \nonumber \\ \!\!\!\!\!\!&&\!\!\!\!\!\! 
\left \langle \frac{\mbox{Tr}_k f_2 (\hat{S}_{k1}^{\alpha}, \hat{S}_{k2}^{\alpha}, \hat{S}_{k3}^{\alpha}, \hat{\sigma}_{k1}^{z}, \hat{\sigma}_{k2}^{z}, \hat{\sigma}_{k3}^{z}) \exp(- \beta \hat{\cal H}_k)}{\mbox{Tr}_k \exp(- \beta \hat{\cal H}_k)} \right \rangle\!. \label{si2} 
\end{eqnarray}
The exact spin identity (\ref{si2}) allows us to express, after straightforward but a little bit tedious calculation, the spontaneous magnetization of the 
Heisenberg spins $m_{\rm H}$ in terms of the triplet correlation $t_{\rm IM} \equiv \langle \hat{\sigma}_{k1}^{z} \hat{\sigma}_{k2}^{z} \hat{\sigma}_{k3}^{z} \rangle_{\rm IM}$ between three Ising spins from the $k$th cluster Hamiltonian and the formerly derived the single-site magnetization $m_{\rm IM}$ of the Ising spins   
\begin{eqnarray}
m_{\rm H} \!\!\!&\equiv&\!\!\! \langle \hat{S}_{k}^{z} \rangle \!=\! \frac{m_{\rm IM}}{2} \!\! \left( \frac{Q_1}{V_1} \!+\! \frac{Q_2}{V_2} \right)
\!+\! \frac{2 t_{\rm IM}}{3} \!\! \left( \frac{Q_1}{V_1} \!-\! 3 \frac{Q_2}{V_2} \right)\!\!, 
\label{mh} 
\end{eqnarray}
and two newly defined functions $Q_1$ and $Q_2$ 
\begin{eqnarray}
Q_1 \!\!\!&=&\!\!\! 3 \exp \left( \frac{3}{4} \beta J_{\rm H} \right) \sinh \left( \frac{3}{2} \beta J_{\rm I} \right) \nonumber \\
\!\!\!&+&\!\!\! \exp \left[ -\frac{1}{4} \beta J_{\rm H}(1 - 4\Delta) \right] \sinh \left( \frac{1}{2} \beta J_{\rm I} \right)  \nonumber \\
\!\!\!&+&\!\!\! 2 \exp \left[ -\frac{1}{4} \beta J_{\rm H}(1 + 2 \Delta) \right] \sinh \left( \frac{1}{2}\beta J_{\rm I} \right), \nonumber \\
Q_2 \!\!\!&=&\!\!\! 3 \exp \left( \frac{3}{4}\beta J_{\rm H} \right) \sinh \left( \frac{1}{2} \beta J_{\rm I} \right) \nonumber \\
\!\!\!&+&\!\!\! \exp \left[-\frac{1}{4} \beta J_{\rm H}(1 + 2\Delta)\right] \sinh \left( \frac{1}{2}\beta J_{\rm I} \right) \nonumber \\
\!\!\!&+&\!\!\! \exp \left[ -\frac{1}{4} \beta J_{\rm H}(1 - \Delta) \right] \cosh \left( \frac{1}{2}\beta Q^{+} \right) \nonumber \\
\!\!\!&-&\!\!\! \exp \left[- \frac{1}{4}\beta J_{\rm H}(1 - \Delta)\right] \cosh \left( \frac{1}{2}\beta Q^{-} \right).
\label{q}
\end{eqnarray}
To complete our exact calculation of the single-site magnetization of the Heisenberg spins $m_{\rm H}$ it is now sufficient to find the triplet correlation 
$t_{\rm IM} \equiv \langle \hat{\sigma}_{k1}^{z} \hat{\sigma}_{k2}^{z} \hat{\sigma}_{k3}^{z} \rangle_{\rm IM}$ of the spin-1/2 Ising model 
on the triangular lattice, which has been rigorously calculated by Baxter and Choy \cite{baxt89}  
\begin{eqnarray}
t_{\rm IM} = \frac{m_{\rm IM}}{4} \!\! \left[ 1 + 2 \frac{y - 2 z^2 + 1 - \sqrt{(y+3)(y-1)}}{y + z^2 - 2} \right]\!\!, 
\label{tim} 
\end{eqnarray} 
where $y = \exp(\gamma \beta J_{\rm eff})$ and $z = \exp(- \gamma \beta J_{\rm eff}/2)$ as before. Substituting the effective nearest-neighbor interaction 
$\gamma J_{\rm eff}$ into the parameters $y$ and $z$ entering the formula (\ref{tim}) one obtains the required closed-form expression for the triplet correlation 
of the Ising spins 
\begin{eqnarray}
t_{\rm IM} = \frac{m_{\rm IM}}{4} \Biggl[ 1 \!\!\!&+&\!\!\! 2 \frac{V_1^{2 \gamma} - 2 V_2^{2 \gamma} + V_1^{\gamma} V_2^{\gamma}}{(V_1^{\gamma} - V_2^{\gamma})^2} \nonumber \\
                     \!\!\!&-&\!\!\! 2 V_1^{\gamma} \frac{\sqrt{(V_1^{\gamma}+3V_2^{\gamma})(V_1^{\gamma}-V_2^{\gamma})}}{(V_1^{\gamma} - V_2^{\gamma})^2}\Biggr]\!. 
\label{ti} 
\end{eqnarray}

Next, let us make a few comments on a critical behavior of the spin-1/2 Ising-Heisenberg model on the TIT lattices. It should be noted here that 
the critical behavior is always accompanied with a non-analytic behavior of the partition function or some of its higher-order temperature derivative. However, 
it may be easily understood from Eqs. (\ref{3})-(\ref{par}) that the mapping parameter $A$ as well as any temperature derivative of it is a smooth continuous 
function and therefore, it must not cause a critical behavior of the spin-1/2 Ising-Heisenberg model. Bearing this in mind, the spin-1/2 Ising-Heisenberg model 
on the TIT lattices becomes critical if and only if the partition function of corresponding spin-1/2 Ising model on the triangular lattice becomes critical as well. 
Accordingly, the critical condition allocating critical points of the spin-1/2 Ising-Heisenberg model on the TIT lattices can readily be obtained from a comparison 
of the effective coupling of the corresponding spin-1/2 Ising model on the triangular lattice with its critical value 
\begin{eqnarray}
\gamma \beta_{\rm c} J_{\rm eff} = \ln 3 \quad  \Leftrightarrow \quad V_1^{\gamma} (\beta_{\rm c}) = 3 V_2^{\gamma} (\beta_{\rm c}).
\label{CritCond}
\end{eqnarray}
Here, it is emphasized that the inverse critical temperature $\beta_{\rm c} = 1/(k_{\rm B} T_{\rm c})$ enters the parameters $V_1$ and $V_2$ given by Eqs. (\ref{3}) and (\ref{4}) instead of $\beta$. It is quite evident from the critical condition (\ref{CritCond}) that the spin-1/2 Ising-Heisenberg model on the TIT lattice is spontaneously long-range ordered whenever the effective coupling is greater than the critical value $\gamma \beta J_{\rm eff} > \ln 3$, otherwise it becomes disordered for $\gamma \beta J_{\rm eff} < \ln 3$. Thus, an intersection of the effective coupling $\gamma \beta J_{\rm eff}$ with its critical value provides a feasible criterion for obtaining the critical points of the spin-1/2 Ising-Heisenberg model on both TIT lattices, which will be thoroughly discussed in Section~\ref{sec:result}B.

Last but not least, let us make a few remarks about a possible extension of our rigorous mapping procedure to a more general case of the spin-1/2 Ising-Heisenberg model on the TIT lattices in a non-zero external magnetic field. While the star-triangle transformation can be rather straightforwardly adapted in order to account for the external magnetic field,\cite{stre10} it will consequently establish a rigorous mapping correspondence with the equivalent spin-1/2 Ising model on a triangular lattice including the effective nearest-neighbor interaction along its edges, the effective triplet interaction within its faces and the effective field acting on its sites. Although the latter classical Ising model is in general not exactly tractable at finite temperatures, one still may take advantage of some precise numerical method (e.g. Monte Carlo simulation) to examine the equivalent purely classical Ising model in an attempt to extract the relevant magnetic behavior of the Ising-Heisenberg model from this exact mapping correspondence. In our follow-up work we have therefore restricted our attention to a rigorous analysis of the zero-temperature magnetization process of the Ising-Heisenberg model on the two considered TIT lattices, which is also compared with the relevant magnetization process of the analogous but purely quantum Heisenberg model treated by means of the exact numerical diagonalization.\cite{cisa13} A remarkable coincidence between the zero-temperature magnetization processes of the Ising-Heisenberg and Heisenberg models displaying the same sequence of intermediate magnetization plateaus proves the usefulness of the simplified Ising-Heisenberg model in providing guidance on a more complex magnetic behavior of the quantum Heisenberg model.\cite{cisa13,ohan12}

\section{Results and discussion}
\label{sec:result}

In this section, let us proceed to a discussion of the most interesting results obtained for the ground state and finite-temperature properties of 
the spin-1/2 Ising-Heisenberg model on two considered TIT lattices. First, it is worth mentioning that all the results derived in the foregoing section are valid 
regardless of whether the interaction constants are assumed to be ferromagnetic or antiferromagnetic. While the change in character of the Heisenberg 
interaction has a profound effect upon the magnetic behavior of the spin-1/2 Ising-Heisenberg model on the TIT lattices, the respective change 
in the Ising interaction $J_{\rm I} \to - J_{\rm I}$ merely causes a rather trivial spin reversal $\sigma_i^z \to -\sigma_i^z$ of all the Ising spins. 
Therefore, we will henceforth restrict our attention only to the particular case of the model with the ferromagnetic Ising interaction $J_{\rm I} > 0$, 
while the respective behavior of the model with the antiferromagnetic Ising interaction $J_{\rm I} < 0$ can simply be deduced from the results to be presented. 
Let us finally mention that the Ising interaction will be subsequently used as the energy unit when defining two dimensionless parameters: the dimensionless 
temperature $ k_{\rm B} T /J_{\rm I}$ and a relative strength of the Heisenberg interaction with respect to the Ising one $J_{\rm H}/J_{\rm I}$.

\subsection{Ground state}

\begin{figure}[t] 
\vspace{-1.2cm}
\includegraphics[width=8cm]{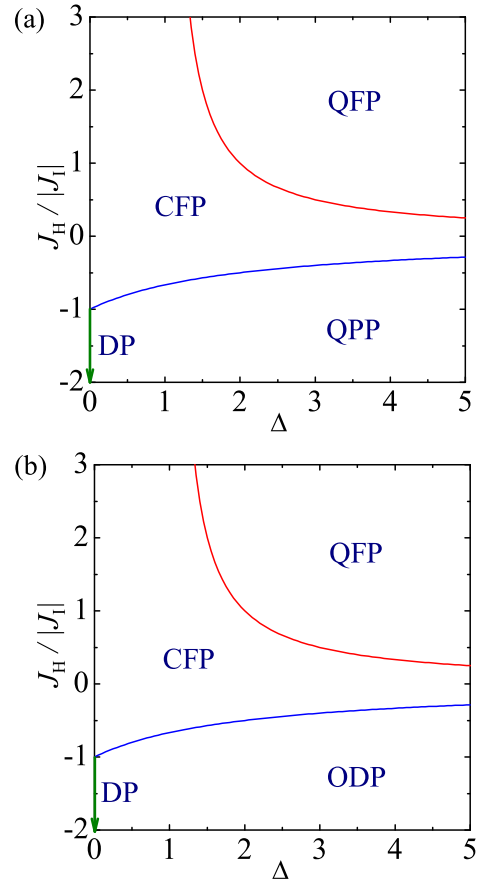}
\vspace{-1.1cm}
\caption{\small (Color online) Ground-state phase-diagrams of the spin-1/2 Ising-Heisenberg model on the two TIT lattices 
shown in Fig. \ref{fig1}(a) and (b). For a detailed description of the phases see the text.}
\label{fig2}
\end{figure}

Let us examine first the ground state of the spin-1/2 Ising-Heisenberg model on two geometrically related TIT lattices under consideration. Because of the commuting character of different cluster Hamiltonians, the overall ground-state spin arrangement is simply given by a tensor product over the lowest-energy eigenstates of the cluster Hamiltonian (\ref{1}). If the lowest-energy eigenvalue of the cluster Hamiltonian (\ref{1}) is degenerate, however, there does not exist a unique ground state but the highly degenerate ground-state manifold whose degeneracy might represent rather complex counting problem. Typical ground-state phase diagrams in the $\Delta - J_{\rm H} / J_{\rm I}$ plane including all possible ground states are depicted in Fig.~\ref{fig2}(a)-(b) for both investigated TIT lattices. Although the ground-state boundaries are identical for both the investigated models, there is a fundamental difference in the character of spin arrangements of some ground-state phases. As one can see from Fig.~\ref{fig2}(a), the Ising-Heisenberg model on the first TIT lattice from Fig.~\ref{fig1}(a) displays two spontaneously long-range ordered and one disordered ground state, more specifically, the classical ferromagnetic phase (CFP), the quantum ferromagnetic phase (QFP), and the disordered quantum paramagnetic phase (QPP). All three phases are mutually separated by two lines of discontinuous (first-order) phase transitions given by the conditions
\begin{eqnarray}
&& \mbox{CFP-QFP:} \qquad \qquad \frac{J_{\rm H}}{J_{\rm I}} = \frac{1}{\Delta - 1}, \,\, (\mbox{for} \, \Delta > 1)
\label{PBa} \\
&& \mbox{CFP-QPP(ODP):}  \quad \frac{J_{\rm H}}{J_{\rm I}} = - \frac{2}{2 + \Delta},
\label{PBb} 
\end{eqnarray}
which were obtained by comparing the energy of the lowest-energy eigenstates constituting the respective ground states.
As could be expected, CFP with a perfect alignment of all the Ising as well as Heisenberg spins 
\begin{eqnarray}
\vert \mbox{CFP} \rangle = \prod_{i=1}^{N} \left  \vert \uparrow \right \rangle_{\!\sigma_{i}^z} 
                           \prod_{k=1}^{N \gamma} \left \vert \uparrow \uparrow \uparrow \right \rangle_{\!S_{k1}^z,S_{k2}^z,S_{k3}^z},
\label{cfp} 
\end{eqnarray}
dominates in the prevailing region of the parameter space with the ferromagnetic Heisenberg interaction $J_{\rm H} > 0$. If the Heisenberg intra-trimer interaction has a strong easy-plane anisotropy, however, one may also encounter a more striking spontaneous long-range order
\begin{eqnarray}
\vert \mbox{QFP} \rangle \!=\! \prod_{i=1}^{N} \! \left  \vert \uparrow \right \rangle_{\!\sigma_{i}^z} \!\! \prod_{k=1}^{N \gamma} \!\! \frac{1}{\sqrt{3}} \! 
\left( \left \vert \uparrow \uparrow \downarrow \right \rangle \!+\! \left \vert \uparrow \downarrow \uparrow \right \rangle 
\!+\! \left \vert \downarrow \uparrow \uparrow \right \rangle \right)_{\!S_{k1}^z,S_{k2}^z,S_{k3}^z}\!\!\!,
\label{qfp} 
\end{eqnarray} 
in which a perfect alignment of all the Ising spins is accompanied with a symmetric quantum superposition of three up-up-down spin states of the Heisenberg trimers. 
Accordingly, the spontaneous magnetization of the Heisenberg spins undergoes a quantum reduction of  the magnetization to one third of the saturation magnetization 
and hence, we will refer to QFP as to the quantum ferromagnetic phase in view of a parallel orientation of both sublattice magnetizations of the Ising and Heisenberg spins, respectively. An origin of this unconventional spontaneous long-range ordering can be related to a competition between two different but ferromagnetic interactions, 
namely, the easy-axis Ising and easy-plane Heisenberg ($\Delta>1$) interaction. 

\begin{figure}[t] 
\vspace{-0.1cm}
\includegraphics[width=8.5cm]{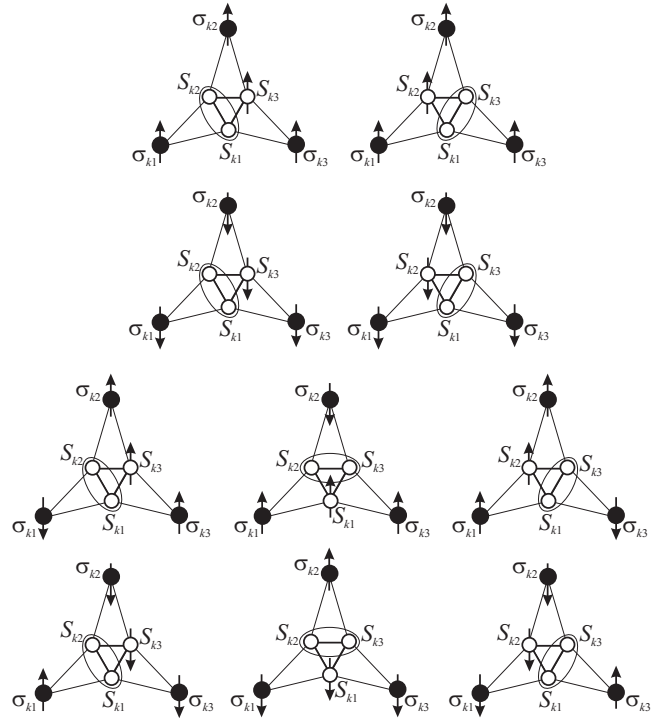}
\vspace{-0.1cm}
\caption{\small The lowest-energy eigenstates of the cluster Hamiltonian (\ref{1}), which constitute the ground-state manifold for the QPP and ODP phases. 
The spin polarization is given by an arrow and an oval denotes a singlet-dimer state.}
\label{fig3a}
\end{figure}

Last but not least, the parameter region with the antiferromagnetic Heisenberg interaction is dominated by QPP without any spontaneous long-range order. 
The disordered phase occurs on behalf of a spin frustration of the Heisenberg trimers, which are incapable to simultaneously satisfy specific requirements of the antiferromagnetic intra-trimer interaction between the Heisenberg spins. If the antiferromagnetic intra-trimer interaction is stronger than the threshold value (27), then, the Heisenberg trimers have a tendency to form one singlet dimer and one unbound spin. The Ising spin coupled to  two Heisenberg spins creating singlet pair will be consequently decoupled from the rest of the respective spin cluster and in this way, the spin frustration of the Heisenberg trimers is also mediated to the Ising spins. Despite the spin frustration, the Ising spins still have a tendency of showing ferromagnetic short-range order as evidenced by the zero-temperature limit of the effective coupling $\lim_{T \to 0} \gamma \beta J_{\rm eff} = \ln 2$ $(\gamma = 1)$ in the corresponding spin-1/2 Ising model on the triangular lattice implying that $\langle \sigma_{i}^z \sigma_{j}^z \rangle_{n.n.} \simeq 0.06564$. The ferromagnetic short-range order of the Ising spins is related to tenfold degeneracy of the lowest-energy eigenstate of the cluster Hamiltonian (3), whereas four out of ten lowest-energy eigenstates of the Heisenberg trimer occur as long as the three enclosing Ising spins are equally aligned and another six provided that one out of three Ising spins points in opposite with respect to the other two (see Fig.~\ref{fig3a} for pictorial representation of all lowest-energy eigenstates)  
\begin{eqnarray} 
\begin{array}{ll} 
\vert \mbox{QPP} \rangle \\ 
\vert \mbox{ODP} \rangle 
\end{array} \!\!\!\! 
= \Biggl \{ 
\begin{array}{llll} 
\left \vert \uparrow \uparrow \uparrow \right \rangle_{\!\sigma_{k1}^z,\sigma_{k2}^z,\sigma_{k3}^z} \frac{1}{\sqrt{2}} \! 
\left( \left \vert \uparrow \downarrow \uparrow \right \rangle \!-\! \left \vert \downarrow \uparrow \uparrow \right \rangle \right)_{S_{k1}^z, S_{k2}^z, S_{k3}^z} \\ 
\left \vert \uparrow \uparrow \uparrow \right \rangle_{\!\sigma_{k1}^z,\sigma_{k2}^z,\sigma_{k3}^z} \frac{1}{\sqrt{2}} \! 
\left( \left \vert \uparrow \uparrow \downarrow \right \rangle \!-\! \left \vert \downarrow \uparrow \uparrow \right \rangle \right)_{S_{k1}^z, S_{k2}^z, S_{k3}^z} \\ 
\left \vert \downarrow \downarrow \downarrow \right \rangle_{\!\sigma_{k1}^z,\sigma_{k2}^z,\sigma_{k3}^z} \frac{1}{\sqrt{2}} \! 
\left( \left \vert \uparrow \downarrow \downarrow \right \rangle \!-\! \left \vert \downarrow \uparrow \downarrow \right \rangle \right)_{S_{k1}^z, S_{k2}^z, S_{k3}^z} \\ 
\left \vert \downarrow \downarrow \downarrow \right \rangle_{\!\sigma_{k1}^z,\sigma_{k2}^z,\sigma_{k3}^z} \frac{1}{\sqrt{2}} \! 
\left( \left \vert \uparrow \downarrow \downarrow \right \rangle \!-\! \left \vert \downarrow \downarrow \uparrow \right \rangle \right)_{S_{k1}^z, S_{k2}^z, S_{k3}^z} \\ 
\left \vert \downarrow \uparrow \uparrow \right \rangle_{\!\sigma_{k1}^z,\sigma_{k2}^z,\sigma_{k3}^z} \frac{1}{\sqrt{2}} \! 
\left( \left \vert \uparrow \downarrow \uparrow \right \rangle \!-\! \left \vert \downarrow \uparrow \uparrow \right \rangle \right)_{S_{k1}^z, S_{k2}^z, S_{k3}^z} \\ 
\left \vert \uparrow \downarrow \uparrow \right \rangle_{\!\sigma_{k1}^z,\sigma_{k2}^z,\sigma_{k3}^z} \frac{1}{\sqrt{2}} \! 
\left( \left \vert \uparrow \uparrow \downarrow \right \rangle \!-\! \left \vert \uparrow \downarrow \uparrow \right \rangle \right)_{S_{k1}^z, S_{k2}^z, S_{k3}^z} \\ 
\left \vert \uparrow \uparrow \downarrow \right \rangle_{\!\sigma_{k1}^z,\sigma_{k2}^z,\sigma_{k3}^z} \frac{1}{\sqrt{2}} \! 
\left( \left \vert \uparrow \uparrow \downarrow \right \rangle \!-\! \left \vert \downarrow \uparrow \uparrow \right \rangle \right)_{S_{k1}^z, S_{k2}^z, S_{k3}^z} \\ 
\left \vert \uparrow \downarrow \downarrow \right \rangle_{\!\sigma_{k1}^z,\sigma_{k2}^z,\sigma_{k3}^z} \frac{1}{\sqrt{2}} \! 
\left( \left \vert \uparrow \downarrow \downarrow \right \rangle \!-\! \left \vert \downarrow \uparrow \downarrow \right \rangle \right)_{S_{k1}^z, S_{k2}^z, S_{k3}^z} \\ 
\left \vert \downarrow \uparrow \downarrow \right \rangle_{\!\sigma_{k1}^z,\sigma_{k2}^z,\sigma_{k3}^z} \frac{1}{\sqrt{2}} \! 
\left( \left \vert \downarrow \uparrow \downarrow \right \rangle \!-\! \left \vert \downarrow \downarrow \uparrow \right \rangle \right)_{S_{k1}^z, S_{k2}^z, S_{k3}^z} \\ 
\left \vert \downarrow \downarrow \uparrow \right \rangle_{\!\sigma_{k1}^z,\sigma_{k2}^z,\sigma_{k3}^z} \frac{1}{\sqrt{2}} \! 
\left( \left \vert \uparrow \downarrow \downarrow \right \rangle \!-\! \left \vert \downarrow \downarrow \uparrow \right \rangle \right)_{S_{k1}^z, S_{k2}^z, S_{k3}^z}. 
\end{array} \nonumber \\ 
\label{dp} 
\end{eqnarray} 
If the three enclosing Ising spins are equally aligned (say all three Ising spins are pointing up), one actually finds just the two linearly independent lowest-energy eigenstates composed of one singlet dimer and one Heisenberg spin polarized by the two surrounding Ising spins into their direction (see the first four eigenstates in Eq.~(\ref{dp}) schematically drawn in the first two rows of Fig.~\ref{fig3a}). On the other hand, there exists just one lowest-energy eigenstate of the Heisenberg trimer being composed of the singlet pair and one polarized Heisenberg spin on assumption that one Ising spin is pointing in opposite with respect to the other two. Under this condition, 
the polarized Heisenberg spin is always connected to the two equally aligned Ising spins and the singlet pair decouples the third unequally oriented Ising spin from the remaining spins of the relevant spin cluster (see the last six eigenstates in Eq.~(\ref{dp}) schematically illustrated in the last two rows of Fig.~\ref{fig3a}). Altogether, 
the ground-state manifold inherent to the QPP will be characterized by a rather huge but non-trivial macroscopic degeneracy, which will be explored in a more detail in Section~\ref{sec:result}D by the analysis of entropy.

\begin{figure}[t] 
\vspace{-0.1cm}
\includegraphics[width=8.5cm]{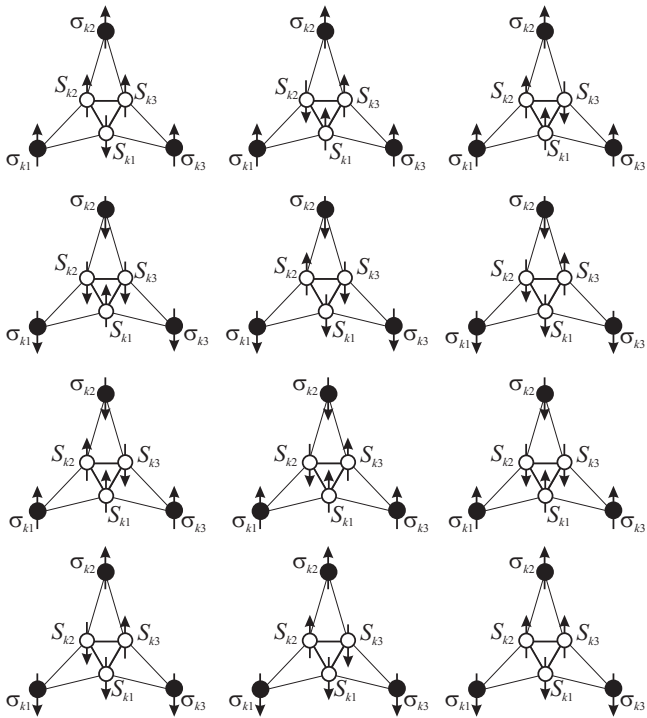}
\vspace{-0.1cm}
\caption{\small The lowest-energy spin configurations, which constitute the ground-state manifold for the DP phase to emerge in the Ising limit $\Delta = 0$. Another twelve lowest-energy spin configurations can be obtained from the spin configurations drawn in the last two lines by rotating them around the threefold C$_3$ symmetry axis.}
\label{fig3b}
\end{figure}

The other disordered phase (DP) is just a special limiting case of QPP for the Ising limit $\Delta=0$, but it deserves a special mention since there appears a dramatic change 
in the character of the lowest-energy eigenstates that constitute the ground-state manifold due to a complete lack of local quantum fluctuations. It can be easily proved that there exist just three different lowest-energy spin configurations on the smaller inner triangles for any given configuration of the three enclosing Ising spins (see Fig.~\ref{fig3b} for the pictorial representation of the lowest-energy configurations). Accordingly, the overall degeneracy of the ground-state manifold becomes for the DP phase a quite simple counting problem $\Omega = 2^{N} 3 ^{\gamma N} = (2 \times 3 ^{\gamma})^{N}$, because the Ising spins residing on the sites of the larger triangles contribute to the overall degeneracy by the factor $2^N$ and the additional factor $3^{\gamma N}$ is related to the three-fold degeneracy of the lowest-energy spin configurations on the smaller inner triangles. Obviously, the similar procedure cannot be used for finding out the degeneracy of the ground-state manifold for the other disordered QPP of the more general Ising-Heisenberg model, since there emerge either two or one lowest-energy eigenstate depending on whether the three enclosing Ising spins 
are equally aligned or not.

As far as the ground state of the spin-1/2 Ising-Heisenberg model on the other TIT lattice (Fig.~\ref{fig1}(b)) is concerned, the ground-state boundaries and the respective spin arrangements of two previously reported spontaneously long-range ordered phases CFP and QFP remain unchanged. However, it surprisingly turns out that the Ising-Heisenberg model on the second TIT lattice displays even in the highly frustrated region a quite peculiar spontaneous long-range order instead of the disordered QPP for arbitrary but non-zero anisotropy parameter $\Delta \neq 0$ (note that the character of the other disordered state DP to emerge for $\Delta = 0$ remains unchanged). The existence of this unconventional spontaneous ordering could be ascribed to the \textit{quantum order-from-disorder} effect, which  partially lifts the ground-state degeneracy of QPP due to a strengthening of local quantum fluctuations invoked by a higher number of the Heisenberg trimers in the second TIT lattice shown in Fig.~\ref{fig1}(b). The quantum order-from-disorder effect generally acts against the spin frustration and it may thus cause an appearance of the unusual partially ordered and partially disordered phase (ODP) with a spontaneously broken Z$_2$ symmetry. Note furthermore that typical spin arrangements to emerge in ODP are completely the same as specified by Eq. (\ref{dp}) for QPP and schematically illustrated in Fig.~\ref{fig3a}. The remarkable spontaneous order of the ODP will be convincingly evidenced in Sections \ref{sec:result}B and \ref{sec:result}C by a more comprehensive analysis of the critical behavior and the order parameter. However, the spontaneous order of ODP could also be evidenced by the effective coupling of the corresponding spin-1/2 Ising model on the triangular lattice tending towards the low-temperature asymptotic limit $\lim_{T \to 0} \gamma \beta J_{\rm eff} = \ln 4$ for 
$\gamma = 2$, which is evidently above the critical value $\gamma \beta_{\rm c} J_{\rm eff} = \ln 3$ and is thus consistent with the spontaneously long-range ordered ground state. It should be pointed out, moreover, that the finite zero-temperature limit of the effective coupling also serves in evidence of an imperfect spontaneous long-range ordering to emerge in ODP, which is also confirmed by non-zero albeit not fully saturated zero-temperature values of the spontaneous magnetizations of the Ising and Heisenberg spins $m_{\rm I} = \frac{1}{2} (\frac{125}{189})^{\frac{1}{8}} \simeq 0.47482$ and $m_{\rm H} = (\frac{2 \sqrt{21}}{27}  - \frac{1}{6}) (\frac{125}{189})^{\frac{1}{8}} \simeq 0.16408$ obtained from Eqs.~(\ref{mi}) and (\ref{mh}), respectively. Altogether, it could be concluded that the local quantum fluctuations are responsible in ODP not only for the quantum reduction of the magnetic moment of the Heisenberg spins, but more strikingly, they also indirectly cause a small but non-zero (cca. 5\%) quantum reduction of the magnetic moment of the classical Ising spins as well.

\begin{figure}[t]
\vspace{-0.3cm} 
\includegraphics[width=6.5cm]{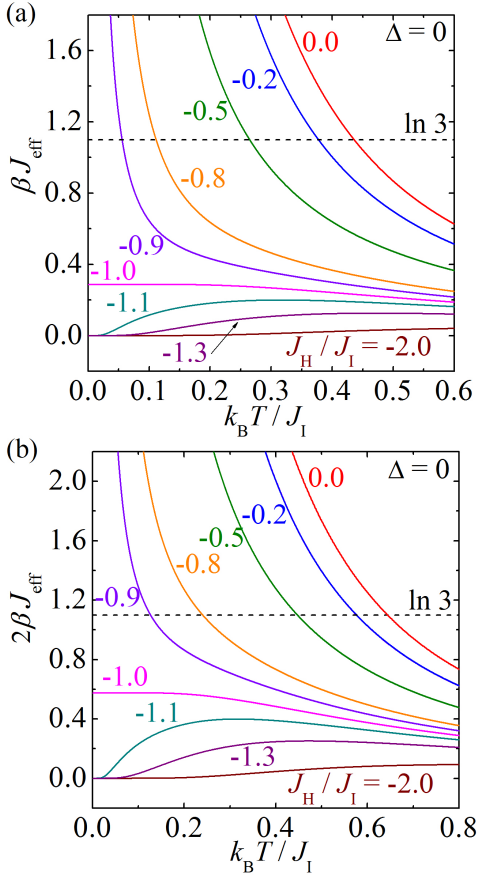}
\vspace{-0.4cm}
\caption{\small (Color online) Temperature dependences of the effective nearest-neighbor interaction $\gamma \beta J_{\rm eff}$ of the corresponding spin-1/2 Ising model on the 
triangular lattice for the specific choice of the anisotropy parameter $\Delta = 0$ and various values of the interaction ratio $J_{\rm H}/J_{\rm I}$. Broken lines show the critical value of the effective coupling above (below) which the spin system becomes spontaneously ordered (disordered).}
\label{fig4}
\end{figure}

\subsection{Critical behavior}

\begin{figure}[t]
\vspace{-0.3cm} 
\includegraphics[width=6.5cm]{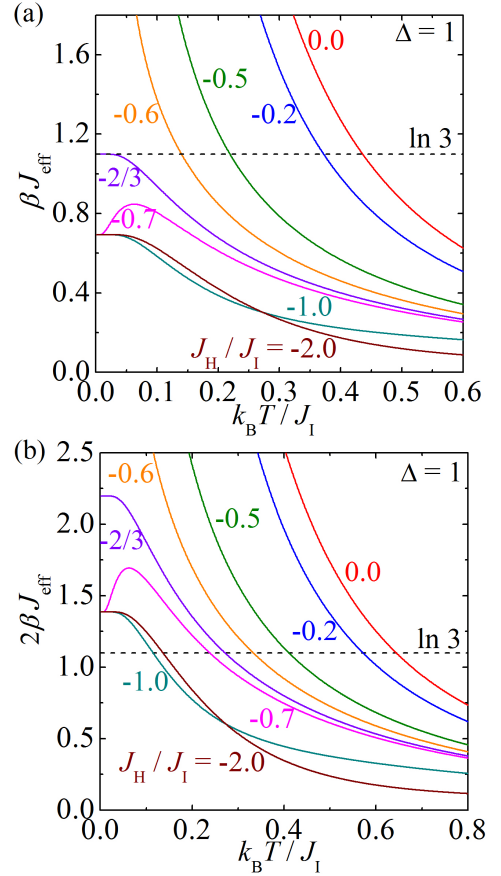}
\vspace{-0.4cm}
\caption{\small (Color online) Temperature dependences of the effective nearest-neighbor interaction $\gamma \beta J_{\rm eff}$ of the corresponding spin-1/2 Ising model on the 
triangular lattice for the specific choice of the anisotropy parameter $\Delta = 1$ and various values of the interaction ratio $J_{\rm H}/J_{\rm I}$. Broken lines show the critical value of the effective coupling above (below) which the spin system becomes spontaneously ordered (disordered).}
\label{fig5}
\end{figure}

Before proceeding to a detailed discussion of the critical behavior, let us explore in detail temperature dependences of the effective interaction $\gamma \beta J_{\rm eff}$ of the corresponding spin-1/2 Ising model on the triangular lattice as depicted in Figs.~\ref{fig4} and \ref{fig5} for two TIT lattices at various values of the interaction ratio $J_{\rm H} / J_{\rm I}$ and two different values of the exchange anisotropy $\Delta = 0$ and $1$, respectively. It can be clearly seen from Fig.~\ref{fig4} that the effective coupling in the Ising limit $\Delta = 0$ either diverges for $J_{\rm H}/J_{\rm I} > -1$ or it tends towards zero for $J_{\rm H}/J_{\rm I} < -1$ as temperature goes to zero. In the latter case, the effective coupling is always below its critical value $\gamma \beta_{\rm c} J_{\rm eff} = \ln 3$ and this result proves an existence of the DP in the Ising limit $\Delta = 0$ of both the investigated TIT models for any temperature if $J_{\rm H}/J_{\rm I} \leq -1$. Contrary to this, the effective coupling $\gamma \beta J_{\rm eff}$ either diverges or asymptotically reaches some finite value for arbitrary but non-zero anisotropy parameter $\Delta \neq 0$ as shown in Fig.~\ref{fig5} for one particular choice of $\Delta$. While the effective coupling $\beta J_{\rm eff}$ of the first TIT lattice always remains below its critical value in the highly frustrated regime $J_{\rm H}/J_{\rm I} < -2/(2 + \Delta)$ (Fig.~\ref{fig5}(a)), the twice as large effective coupling $2 \beta J_{\rm eff}$ of the second TIT lattice is strong enough in order to induce the spontaneous long-range order at sufficiently low temperatures even in the highly frustrated region $J_{\rm H}/J_{\rm I} < -2/(2 + \Delta)$ (Fig.~\ref{fig5}(b)). This result thus provides an independent confirmation of the quantum order-from-disorder effect, which arises from the local quantum fluctuations governing the magnetic behavior of the Heisenberg trimers. Finally, it is worthy to recall that an intersection of the effective coupling $\gamma \beta J_{\rm eff}$ with the relevant critical point of the spin-1/2 Ising model on the triangular lattice can be straightforwardly used in order to locate critical points of the spin-1/2 Ising-Heisenberg model on the TIT lattices, which in fact represents the numerical solution of the critical condition (\ref{CritCond}).  

\begin{figure}[t] 
\vspace{-1.1 cm}
\includegraphics[width=8cm]{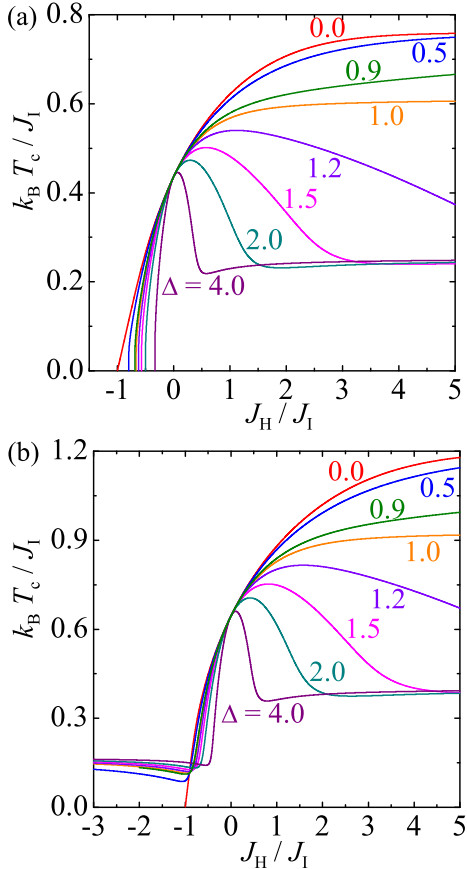}
\vspace{-1.2 cm}
\caption{\small (Color online) Critical temperature $k_{\rm B} T_{\rm c} / J_{\rm I}$ of the spin-1/2 Ising-Heisenberg model on two TIT lattices from Fig. \ref{fig1}(a)-(b) 
as a function of the interaction ratio $J_{\rm H} / J_{\rm I}$ for several values of the exchange anisotropy $\Delta$.}
\label{fig6}
\end{figure}

The critical temperature of two investigated TIT models is plotted in Fig.~\ref{fig6} against a relative strength between the Heisenberg and Ising interaction for several values of the anisotropy parameter $\Delta$. If one considers the particular case of the easy-axis exchange anisotropy $\Delta < 1$, the critical temperature of the first lattice model (Fig.~\ref{fig6}(a)) monotonically decreases with decreasing the ratio $J_{\rm H}/J_{\rm I}$ until it completely vanishes at the ground-state boundary (\ref{PBb}) between CFP and QPP. The highest critical temperature of CFP
\begin{eqnarray}
\displaystyle \lim_{\frac{J_{\rm H}}{J_{\rm I}} \to \infty} \!\! \frac{k_{\rm B} T_{\rm c}}{J_{\rm I}} = \! \left[\ln \! \left(2 + \sqrt{3} \right) \right]^{-1} \!\!\!\!\!\! \simeq 0.75933
\label{tcas1} 
\end{eqnarray} 
can be accordingly acquired in the asymptotic limit $J_{\rm H}/J_{\rm I} \to \infty$. On the other hand, the critical temperature exhibits a more striking non-monotonous dependence when assuming the easy-plane exchange anisotropy $\Delta > 1$, because the sufficiently strong ferromagnetic Heisenberg interaction then favors a presence of unconventional quantum ferromagnetic ordering QFP before the classical CFP one. The critical temperature therefore tends just to one third of the previously reported asymptotic value 
\begin{eqnarray}
\displaystyle \lim_{\frac{J_{\rm H}}{J_{\rm I}} \to \infty} \!\! \frac{k_{\rm B} T_{\rm c}}{J_{\rm I}} = \! \left[3 \ln \! \left(2 + \sqrt{3} \right) \right]^{-1} \!\!\!\!\!\! \simeq 0.25311,
\label{tcas2} 
\end{eqnarray} 
since the magnetic moment of all the Heisenberg spins in QFP is reduced by local quantum fluctuations to one third of their magnetic moment in CFP. Besides, the most conspicuous increase of the critical temperature upon lowering the interaction ratio $J_{\rm H}/J_{\rm I}$ can be detected in a close vicinity of the phase boundary (\ref{PBa}) between QFP and CFP, which is attributable to the increase of the magnetic moment of the Heisenberg spins when passing from QFP towards CFP. As far as the critical behavior of the second TIT lattice (Fig.~\ref{fig6}(b)) is concerned, the critical temperature generally exhibits qualitatively the same dependences on assumption that the Heisenberg interaction 
is ferromagnetic. As a matter of fact, the critical temperature is then shifted towards slightly higher values due to a higher connectivity of the Ising spins in the other TIT lattice and the asymptotic values of the critical temperature in $J_{\rm H}/J_{\rm I} \to \infty$ limit read 
\begin{eqnarray}
\displaystyle \lim_{\frac{J_{\rm H}}{J_{\rm I}} \to \infty} \!\! \frac{k_{\rm B} T_{\rm c}}{J_{\rm I}} = \! \left[\ln \! \left( \frac{1 + \sqrt{3} + \sqrt{2 \sqrt{3}}}{2} \right) \right]^{-1} \!\!\!\!\!\! \simeq 1.20273
\label{tcas3} 
\end{eqnarray} 
for CFP if $\Delta<1$ and
\begin{eqnarray}
\displaystyle \lim_{\frac{J_{\rm H}}{J_{\rm I}} \to \infty} \!\! \frac{k_{\rm B} T_{\rm c}}{J_{\rm I}} =  \!
\left[3 \ln \! \left( \frac{1 + \sqrt{3} + \sqrt{2 \sqrt{3}}}{2} \right) \right]^{-1} \!\!\!\!\!\! \simeq 0.40091
\label{tcas4} 
\end{eqnarray} 
for QFP if $\Delta>1$. The most fundamental difference in the critical frontiers of two investigated TIT models can be thus found in the parameter region with a strong (negative) antiferromagnetic Heisenberg interaction, where the critical lines of the latter model do not vanish but they tend towards some non-zero value after passing through a global minimum in a proximity of the phase transition (\ref{PBb}) between CFP and ODP phases. It could be concluded that the presented exact results for the critical boundaries provide another independent confirmation for an existence of the striking spontaneous long-range order ODP, which emerges in the latter TIT model in spite of the high spin frustration.  
 
\subsection{Spontaneous magnetization}

\begin{figure} 
\vspace{-1.1cm}
\includegraphics[width=8cm]{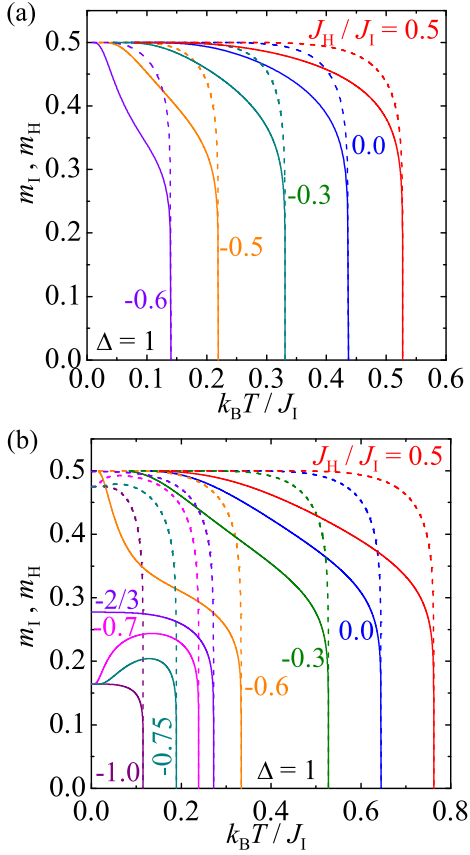}
\vspace{-1.2cm}
\caption{\small (Color online) Temperature dependences of the spontaneous magnetization $m_{\rm I}$ of the Ising spins (broken lines) and the spontaneous magnetization $m_{\rm H}$ of the Heisenberg spins (solid lines) for the fixed value of the anisotropy parameter $\Delta = 1$ and several values of the interaction ratio $J_{\rm H} / J_{\rm I}$.}
\label{fig7}
\end{figure}

\begin{figure}
\vspace{-1.1cm} 
\includegraphics[width=8cm]{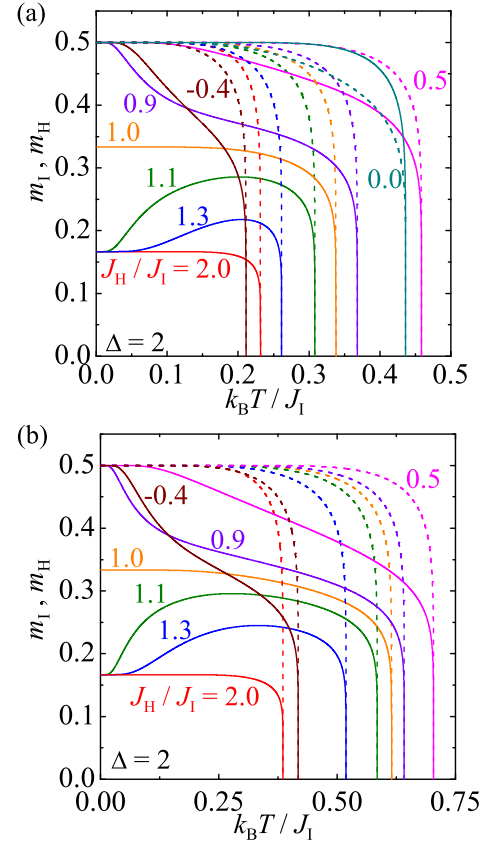}
\vspace{-1.2cm}
\caption{\small (Color online) Temperature dependences of the spontaneous magnetization $m_{\rm I}$ of the Ising spins (broken lines) and the spontaneous magnetization 
$m_{\rm H}$ of the Heisenberg spins (solid lines) for the fixed value of the anisotropy parameter $\Delta = 2$ and several values of the interaction ratio $J_{\rm H}/J_{\rm I}$.}
\label{fig8}
\end{figure}

Now, let us turn our attention to a discussion of typical temperature dependences of the spontaneous magnetization, which will bring insight into the main differences in the thermal behavior of three spontaneously long-range ordered phases CFP, QFP, and ODP. For this purpose, the temperature dependences of both spontaneous sublattice magnetizations are plotted in Figs.~\ref{fig7} and \ref{fig8} for two investigated TIT lattices, two different values of the exchange anisotropy $\Delta$ and several values of the interaction 
ratio $J_{\rm H}/J_{\rm I}$. First, let us take a closer look at thermal variations of the spontaneous magnetizations of the Ising and Heisenberg spins in two TIT lattices 
with the isotropic Heisenberg intra-trimer interaction. It is quite obvious from Fig.~\ref{fig7}(a) that the spontaneous magnetizations of the Ising and Heisenberg spins 
for the first TIT lattice start from their saturated values, which bear evidence of CFP unless the disordered QPP becomes the ground state for $J_{\rm H}/J_{\rm I} < -2/3$. 
In addition, it is also quite clear from Fig.~\ref{fig7}(a) that the spontaneous magnetizations of the Ising and Heisenberg spins exhibit a similar temperature-induced decline, the magnetization of the Heisenberg spins actually shows only a slightly greater temperature-induced downturn than the spontaneous magnetization of the Ising spins even though both sublattice magnetizations tend to zero with the same critical exponent $\beta_{\rm e} = 1/8$ from the standard universality class of the two-dimensional Ising model. It should be nevertheless noted here that the same general trends can also be detected in the respective temperature dependences of the spontaneous magnetizations of the second TIT lattice whenever $J_{\rm H}/J_{\rm I} > -2/3$ (see Fig.~\ref{fig7}(b)). However, the spontaneous magnetizations of the Ising and Heisenberg spins in the latter TIT model start from non-zero but not fully saturated values $m_{\rm I} \simeq 0.47482$ and $m_{\rm H} \simeq 0.16408$ even if $J_{\rm H}/J_{\rm I} < -2/3$, which bear evidence of the unconventional spontaneous ordering ODP basically affected by the local quantum fluctuations. The spontaneous magnetization of the Heisenberg spins reveals the transition between CFP and ODP through a vigorous temperature-induced increase of this sublattice magnetization as long as the interaction ratio $J_{\rm H}/J_{\rm I}$ is selected sufficiently close to but slightly below the ground-state boundary (\ref{PBb}) (see the curves for $J_{\rm H}/J_{\rm I} = -0.7$ and $-0.75$ in Fig.~\ref{fig7}(b)).  

To shed light on another unusual spontaneous long-range ordered phase QFP, Fig.~\ref{fig8} displays temperature variations of the spontaneous magnetization of two investigated TIT lattices for the ferromagnetic Heisenberg interaction with the easy-plane exchange anisotropy $\Delta = 2$. Under this condition, the spontaneous magnetizations of the Ising and Heisenberg spins apparently exhibit according to Fig.~\ref{fig8} qualitatively the same temperature dependences for both investigated TIT lattices. If the relative strength between the Heisenberg and Ising interaction exceeds the boundary value (\ref{PBa}), one actually finds that the initial value of the spontaneous magnetization of the Heisenberg spins is reduced by the local quantum fluctuations to one third of its saturation magnetization while the initial value of the spontaneous magnetization of the Ising spins is still fully saturated. This observation is fully consistent with the ground-state spin arrangement that was attributed to QFP. Beside this, thermal excitations in QFP may give rise to an interesting increase in the spontaneous magnetization of the Heisenberg spins, whereas the observed temperature-induced increase in $m_{\rm H}$ is the greater, the closer the interaction ratio $J_{\rm H}/J_{\rm I}$ is selected to the ground-state boundary (\ref{PBa}). Note furthermore that one may also get the special value of the spontaneous magnetization of the Heisenberg spins $m_{\rm H} = 1/3$ due to the coexistence of the CFP and QFP if the interaction ratio is selected exactly at their ground-state boundary (\ref{PBa}), i.e. $J_{\rm H}/J_{\rm I} = 1$ for the particular case of $\Delta = 2$.

\subsection{Entropy}

\begin{figure}
\vspace{-1.1cm} 
\includegraphics[width=7.8cm]{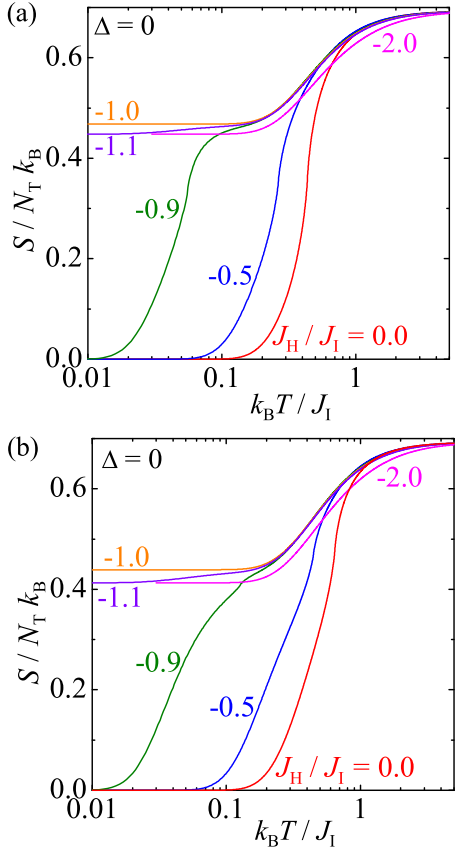}
\vspace{-1.2cm}
\caption{\small (Color online) Semilogarithmic plot for temperature dependences of the entropy per one spin $S / N_{\rm T} k_{\rm B}$ for two considered TIT lattices, the special value of the exchange anisotropy $\Delta = 0$ and several values of the interaction ratio $J_{\rm H} / J_{\rm I}$.}
\label{fig9}
\end{figure}

\begin{figure}
\vspace{-1.1cm} 
\includegraphics[width=7.8cm]{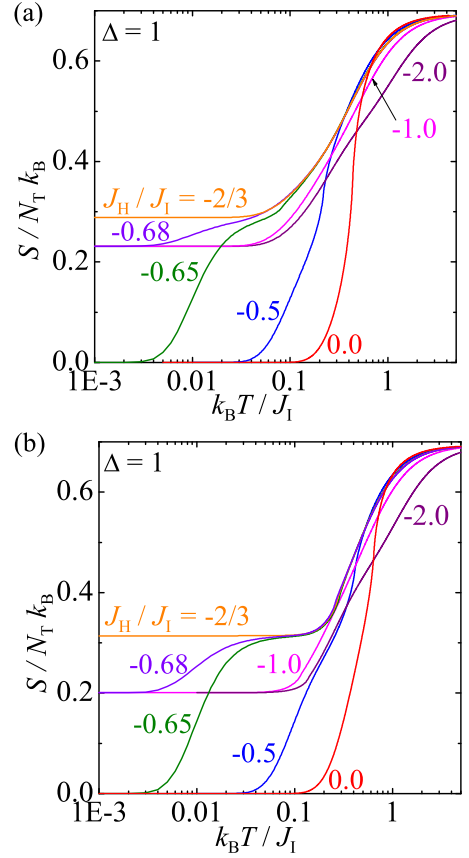}
\vspace{-1.2cm}
\caption{\small (Color online) Semilogarithmic plot for temperature dependences of the entropy per one spin $S / N_{\rm T} k_{\rm B}$ for two considered TIT lattices, the special value of the exchange anisotropy $\Delta = 1$ and several values of the interaction ratio $J_{\rm H} / J_{\rm I}$.}
\label{fig10}
\end{figure}

Next, let us turn to a detailed analysis of the entropy, which enables a deeper insight into the degree of randomness in the disordered states. Figs.~\ref{fig9} and \ref{fig10}
depict temperature variations of the entropy per one spin $S/N_{\rm T} k_{\rm B}$ ($N_{\rm T} = N (1 + 3 \gamma)$ is the total number of all spins) for two different values of the exchange anisotropy and several values of the interaction ratio $J_{\rm H}/J_{\rm I}$. It is quite evident from Fig.~\ref{fig9} that the highly macroscopically degenerate ground state DP develops in the Ising limit $\Delta = 0$ of both investigated lattice models due to the spin frustration, which originates from the sufficiently strong antiferromagnetic intra-trimer interaction $J_{\rm H}/J_{\rm I}<-1$. Consequently, the DP manifests itself through a relatively large residual entropy $S/N_{\rm T} k_{\rm B} = \frac{1}{4} \ln 6 \simeq 0.44794$ and $\frac{1}{7} \ln 18 \simeq 0.41291$, which is in agreement with the calculated degeneracy of the ground-state manifold inherent to the DP in the limiting Ising case (see Section \ref{sec:result}A). On the other hand, arbitrary but non-zero anisotropy parameter $\Delta \neq 0$ is responsible for an onset of the local quantum fluctuations, which lift the macroscopic degeneracy approximately by $50\%$ in the highly frustrated regime of both TIT lattices (see Fig.~\ref{fig10} for $\Delta = 1$). While the residual entropy of the former TIT lattice $S/N_{\rm T} k_{\rm B} \simeq 0.23133$ is still high enough to preserve the disordered nature of QPP, the slightly lower residual entropy of the latter TIT lattice $S/N_{\rm T} k_{\rm B} \simeq 0.20075$ supports an existence of partially ordered and partially disordered state ODP. In spite of the spontaneously broken symmetry, the ODP still retains a relatively high degree of randomness and is quite reminiscent of the fully disordered QPP.

\subsection{Specific heat}

\begin{figure} 
\vspace{-1.1cm}
\includegraphics[width=8cm]{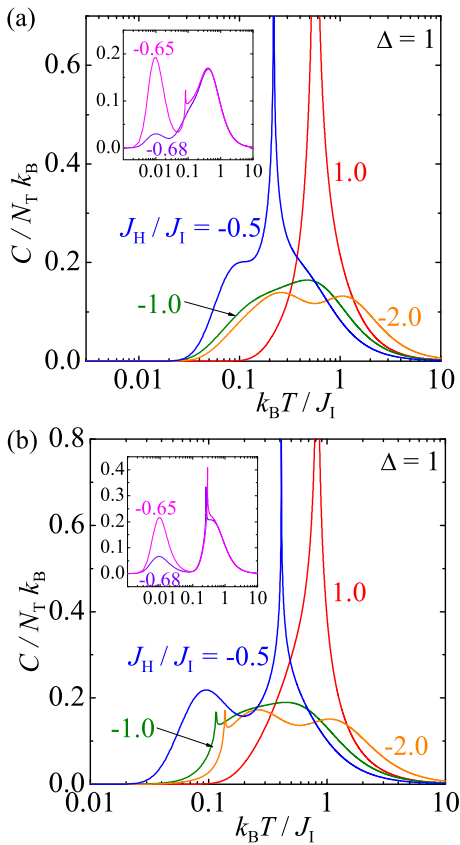}
\vspace{-1.2cm}
\caption{\small (Color online) Semilogarithmic plot for temperature dependences of the specific heat per one spin $C / N_{\rm T} k_{\rm B}$ for two considered TIT lattices, the special value of the exchange anisotropy $\Delta = 1$ and several values of the interaction ratio $J_{\rm H} / J_{\rm I}$.}
\label{fig11}
\end{figure}

\begin{figure}
\vspace{-1.1cm}
\includegraphics[width=8cm]{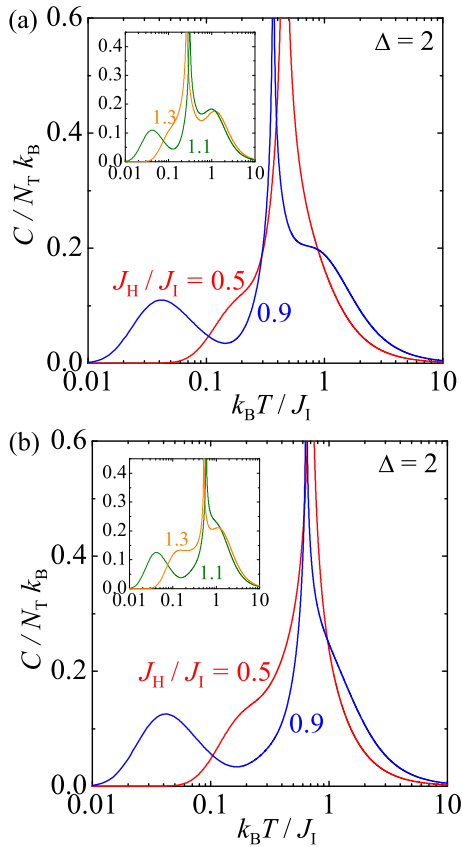}
\vspace{-1.2cm}
\caption{\small (Color online) Semilogarithmic plot for temperature dependences of the specific heat per one spin $C / N_{\rm T} k_{\rm B}$ for two considered TIT lattices, the special value of the exchange anisotropy $\Delta = 2$ and several values of the interaction ratio $J_{\rm H} / J_{\rm I}$.}
\label{fig12}
\end{figure}

Finally, let us discuss the main features of temperature dependences of the zero-field specific heat, which are displayed in a semilogarithmic scale in Figs.~\ref{fig11} and \ref{fig12} for two investigated TIT lattices by selecting two different values of the exchange anisotropy $\Delta$ and various values of the interaction ratio $J_{\rm H}/J_{\rm I}$. Typical thermal variations of the specific heat for the particular case of the isotropic Heisenberg intra-trimer interaction are illustrated in Fig.~\ref{fig11}. As one can see from Fig.~\ref{fig11}(a), the Ising-Heisenberg model on the first TIT lattice exhibits a logarithmic singularity from the standard Ising universality class when considering the ferromagnetic Heisenberg interaction $J_{\rm H}>0$. It is quite evident from this figure, moreover, that a gradual strengthening of the antiferromagnetic intra-trimer interaction causes an appearance of a shoulder superimposed on the low-temperature tail of the specific heat divergence (see the curve for $J_{\rm H}/J_{\rm I} = -0.5$). If the interaction ratio $J_{\rm H}/J_{\rm I}$ is selected sufficiently close but slightly above the ground-state boundary (\ref{PBb}) between CFP and QPP, the marked round maximum at relatively low temperatures is subsequently followed by the logarithmic divergence superimposed on ascending part of another round Schottky-type maximum emerging at higher temperatures (see the curve for $J_{\rm H}/J_{\rm I} = -0.65$ in the inset of Fig.~\ref{fig11}(a)). In agreement with the ground-state and finite-temperature phase diagrams displayed in Figs.~\ref{fig2}(a) and \ref{fig6}(a), the logarithmic singularity completely vanishes from the temperature dependence of the heat capacity with regard to the disordered character of QPP whenever the antiferromagnetic intra-trimer interaction exceeds the ground-state boundary (\ref{PBb}) between CFP and QPP (see the curves for $J_{\rm H}/J_{\rm I} < -2/3$). After passing through this ground-state boundary, the low-temperature round maximum rather steeply diminishes within QPP as the Heisenberg interaction further strengthens (see the curve for $J_{\rm H}/J_{\rm I} = -0.68$ in the inset of Fig.~\ref{fig11}(a)), then it shows a single round maximum of a rather irregular shape for the Heisenberg interaction of a moderate strength (e.g. for $J_{\rm H}/J_{\rm I} = -1$) and finally, the specific heat displays two more or less separated round maxima for a relatively strong Heisenberg interaction (e.g. for $J_{\rm H}/J_{\rm I} = -2$). As far as the specific heat of the second TIT lattice model is concerned (Fig.~\ref{fig11}(b)), it generally shows similar thermal variations except that one still encounters a marked logarithmic singularity inside the highly frustrated region ($J_{\rm H}/J_{\rm I} < -2/3$ for $\Delta=1$), which is occupied by the unconventional partially ordered and partially disordered state ODP rather than the fully disordered QPP.

Last, let us briefly comment on a typical thermal behavior of the heat capacity in a close vicinity of the another possible phase transition between CFP and QFP. For this purpose, Fig.~\ref{fig12} illustrates typical temperature dependences of the specific heat for the particular case of the ferromagnetic Heisenberg interaction with one selected value of the easy-plane exchange anisotropy $\Delta = 2$. According to these plots, the specific heat of both considered TIT lattices exhibits qualitatively the same thermal variations with only a small shift of the logarithmic singularity of the latter model towards higher temperatures. If the interaction parameters are tuned sufficiently close to the ground-state boundary between CFP and QFP ($J_{\rm H}/J_{\rm I} = 1$ for $\Delta=2$), then, one observes a development of the round Schottky-type maximum at relatively low temperatures reflecting spin excitations from CFP to QFP (for $J_{\rm H}/J_{\rm I} < 1$) or vice versa (for $J_{\rm H}/J_{\rm I} > 1$). The closer is the ratio 
$J_{\rm H}/J_{\rm I}$ to the ground-state boundary (\ref{PBa}), the more pronounced the low-temperature Schottky-type maximum can be observed in the relevant thermal dependence. In addition to this rather robust low-temperature round maximum, there also may appear the second high-temperature round maximum upon further increase of a relative strength of the Heisenberg interaction as convincingly evidenced by temperature variations of the specific heat displayed in the inset of Fig.~\ref{fig12} for $J_{\rm H}/J_{\rm I} = 1.3$.    
 
\section{Concluding remarks}
\label{sec:conclusion}

In the present article, the spin-1/2 Ising-Heisenberg model on two different but geometrically related TIT lattices has been exactly solved through the generalized star-triangle transformation establishing a rigorous mapping equivalence with the corresponding spin-1/2 Ising model on a triangular lattice. Within the framework of this rigorous mapping method, we have derived exact analytical results for several basic thermodynamic quantities such as the free and internal energy, spontaneous magnetization, entropy, specific heat, and we have also constructed the ground-state and finite-temperature phase diagrams quite rigorously. It has been demonstrated that the spin-1/2 Ising-Heisenberg model on two TIT lattices exhibit a surprisingly rich magnetic behavior including several unconventional and yet undetected quantum phases without any classical counterpart. 

A mutual competition between two ferromagnetic interactions of basically different character (the easy-axis Ising and easy-plane Heisenberg interactions) is the main cause for the emergence of QFP in which a symmetric quantum superposition of three up-up-down states of the Heisenberg trimers accompanies a perfect alignment of all the Ising spins. Apart from this common feature of both studied lattice models, it has been evidenced that two investigated TIT lattices display a very different spin ordering (QPP versus ODP) in the highly frustrated region owing to the crucial difference in a relative strength of local quantum fluctuations. Among the most remarkable findings one should mention an existence of ODP alone, which is in part spontaneously ordered as evidenced by the singular behavior and power-law decay of several thermodynamic quantities in a close vicinity of the critical point, and in part disordered, as evidenced by the non-zero residual entropy. It actually turns out that the local quantum fluctuations are responsible in ODP not only for a quantum reduction of the spontaneous magnetization of the Heisenberg spins, but they also indirectly cause a quite peculiar quantum reduction of the spontaneous magnetization of the otherwise classical Ising spins. The present work thus provides the first example of exactly solved frustrated spin model with a mutual coexistence of the imperfect spontaneous order and the partial disorder, which still exhibits a non-trivial criticality at finite temperatures. 

\begin{acknowledgments}
The authors would like to thank prof. Fr\'ed\'eric Mila for his valuable comments and insights, which helped us in perceiving the nature of unusual quantum ground states emerging in the highly frustrated regime. This work was accomplished under the financial support of the Sciex fellowship No.~11.056. The financial support provided by the Scientific Grant Agency of Ministry of Education of Slovak Republic under the VEGA Grant No.~1/0234/12 is also gratefully acknowledged.
\end{acknowledgments}

\end{document}